\documentclass[prd,10pt,aps,showpacs,superscriptaddress,twocolumn,amsmath,amssymb,nofootinbib,longbibliography]{revtex4-1}
\usepackage[english]{babel}

\usepackage{graphicx,subfigure,float}
\usepackage[colorlinks=true, allcolors=blue]{hyperref}

\usepackage{orcidlink}

\usepackage{tikz}
\usetikzlibrary{shadings}
\usepackage{xcolor}

\begin{document}

\title{Gravitational Effects on Light Propagation in Linear Magnetoelectrics and Plasmas}

\author{Elda \surname{Guzm\'an-Herrera}\orcidlink{0000-0003-1400-9332}}
\email{elda.guzman@unifei.edu.br}
\affiliation{Instituto de F\'{\i}sica e Qu\'{\i}mica, Universidade Federal de Itajub\'a, \\
Itajub\'a, Minas Gerais 37500-903, Brazil}
\author{Renato \surname{Klippert}\orcidlink{0000-0001-6977-0124}}
\email{klippert@unifei.edu.br}
\affiliation{Instituto de F\'{\i}sica e Qu\'{\i}mica, Universidade Federal de Itajub\'a, \\
Itajub\'a, Minas Gerais 37500-903, Brazil}
\author{Vitorio A. \surname{De Lorenci}\orcidlink{0000-0001-5880-2207}}
\email{delorenci@unifei.edu.br}
\affiliation{Instituto de F\'{\i}sica e Qu\'{\i}mica, Universidade Federal de Itajub\'a, \\
Itajub\'a, Minas Gerais 37500-903, Brazil}
\begin{abstract}
A covariant formalism for electromagnetism is used to investigate the effects of light propagation within certain material systems under the influence of a background gravitational field. Specifically, dielectric media described by linear constitutive relations, as well as magnetized plasma media, such as the magnetosphere of neutron stars, are examined. Estimates for detecting gravitational effects through precise refractive index measurements within these scenarios are provided.
\end{abstract}

\maketitle

\section{Introduction}
The influence of gravity on light propagation in Earth-based optical systems is usually neglected because of its small magnitude. However, as precision metrology advances \cite{2026AdPhN...5a6015Z,2025OptCo.58631829L}, such corrections are becoming measurable and could be fundamental in exploring the foundational limits of physics. Moreover, gravity is dominant in many astrophysical systems, where light propagation is appreciably affected by strong gravitational fields. At the same time, the study of the gravitational contribution to the refractive index is also motivated by the increasing precision of interferometric experiments, as observatories such as LIGO and the future LISA mission \cite{arun2022new} push toward ultra-precision measurements.

The propagation of electromagnetic waves in material systems is mainly determined by the electromagnetic response of the medium. Even in the linear regime, optical materials exhibit a rich phenomenology arising from the coupling between electric and magnetic fields, which led to technological applications \cite{ortega2015multifunctional,shiratsuchi2021magnetoelectric,gupta2022review,eerenstein2006multiferroic}. In particular, magnetoelectric materials have attracted considerable attention because polarization can be induced by magnetic fields and magnetization by electric fields, leading to many optical phenomena \cite{fiebig2005revival,scott2013room}.

On the other hand, plasmas are ionized fluids that exhibit anisotropic dielectric properties under the action of applied electromagnetic fields. They are claimed to be the most abundant form of visible matter in the Universe \cite{chatterjee2022fundamental}. Wave propagation phenomena in such systems have been largely investigated in laboratory experiments \cite{gekelman2016upgraded,boxer2010turbulent,amatucci2003laboratory} and in astrophysical environments \cite{kobialko2024black,fleishman2002birefringence,zhao2017properties}. In compact objects such as pulsars and magnetars, the interplay of intense magnetic fields and strong gravitational fields in the magnetosphere leads to important optical effects, which could be useful for extracting relevant information from these extreme objects, such as their mass, radius, and magnetic field.

An appropriate framework for investigating light propagation in material systems and in a curved spacetime is the covariant formulation of electromagnetism. Maxwell's equations can be written in a manifestly covariant form in terms of the field tensors $F_{\mu\nu}$ and $\mathcal{G}^{\mu\nu}$, which allow the constitutive properties of the medium to be treated independently of coordinate systems and the background spacetime metric \cite{de2023remarks,de2022aspects}. Thus, it is convenient to define the constitutive relations in terms of the fundamental fields $\bf{E}$ and $\bf{B}$ (those appearing in Faraday's induction law), rather than the auxiliary fields $\bf{D}$ and $\bf{H}$ \cite{de2023remarks}.

In the next section, the study of light propagation in a linear material medium is presented using a covariant formalism. The constitutive relations for such a medium are derived from the thermodynamical description of its free-energy density. The covariant form of the Fresnel tensor is also determined, which includes the gravity contribution by means of the background spacetime metric. The analysis is restricted to the lossless linear regime, allowing the gravitational contribution to be compared directly with the intrinsic optical response of the material. 
In Sec. ~\ref{optics_linear}, a specific class of crystalline optical material is chosen, and the corresponding refractive indices in either Earth- or Solar-based systems are derived. The different contributions to refractive indices are discussed, and some estimates are presented. Present-day technology suggests that the contribution of gravity to birefringence effects could be detected using interferometry. Wave propagation in a magnetized plasma described within the magnetohydrodynamics regime in a curved spacetime is investigated in Sec.~\ref {plasmasection}. Two physical environments are discussed: a Tokamak plasma \cite{petrov2006influence,shen2023plasma} and a highly magnetized neutron-star magnetosphere. Phase and group velocities are obtained, and their dependence on gravity is examined in the context of cold and warm plasma regimes. Conclusions are presented in Sec.~\ref{final}, and some considerations about the equivalence of the three- and four-dimensional formalisms for linear optics are addressed in the appendix.

The units of the electric and magnetic fields are $[{\rm E}]={\rm V\, m}^{-1}$, $[{\rm B}]={\rm T}$, while the permittivity, permeability, and magnetoelectric tensors are $[\varepsilon]=[\varepsilon_{0}]$, $[\mu]=[\mu_{0}]$ and $[\alpha \mu_{0}]={\rm s\, m}^{-1}$. 
Einstein's convention for sums is employed throughout. The four-dimensional Levi-Civita symbol $\eta^{\alpha\beta\gamma\delta}$ is a completely antisymmetric rank-4 quantity with $\eta^{0123}=1$. The 4-wave vector is defined as $k_{\mu}=(\omega/c,q_{i})$, with $c$ being the value of the speed of light in vacuum.

\section{Constitutive relations and wave propagation in the Schwarzschild background}
The free-energy density $F$ of an optical crystalline material, at fixed  
volume $V$ and temperature $T$, can be expanded in powers of the 
electromagnetic fields ${\bf E}$ and ${\bf B}$, which are thus assumed to be thermodynamic variables. Since zeroth- and first-order terms do not contribute to the propagation of electromagnetic waves, and higher-order nonlinearities were neglected, only the quadratic contributions in the fields is retained 
\cite{landau2013electrodynamics, de2023remarks, rivera2009short}
\begin{eqnarray*}
    F(E_{i},B_{i})  = -\tfrac{\varepsilon_{0}}{2}\chi^{ij}E_{i}E_{j} -\tfrac{1}{2\mu_{0}}\tilde{\chi}^{ij}B_{i}B_{j} -\alpha^{ij}E_{i}B_{j},
\end{eqnarray*}
where the dimensionless coefficients $\chi^{ij}$ and $\tilde{\chi}^{ij}$ denote the electric and magnetic susceptibilities of the medium, and $\alpha^{ij}$ are the linear magnetoelectric coefficients.
It is of worth mention that this expansion includes only linear optical effects, i.e., the dielectric, magnetic, and magnetoelectric responses to external fields. The magnetoelectric contribution is activated if either an electric or a magnetic field is applied, inducing, respectively, a magnetization or a polarization of the medium, without requiring the background components of both fields to be simultaneously present. Investigations on higher-order magnetoelectric phenomena have also been considered elsewhere \cite{de2023remarks}.

The polarization and magnetization of the medium can be obtained as $P^{i} = -(\partial F/\partial E_{i})$ and $M^{i}  =  -(\partial F/\partial B_{i})$. It is convenient to express these relations in terms of the auxiliary fields $D^{i}=\varepsilon_{0}\delta^{ij}E_{j}+P^{i}$ and $H^{i}=\tfrac{1}{\mu_0}\delta^{ij}B_{j}-M^{i}$, which read
\begin{eqnarray}
    D^{i}& = & P_{S}^{i}+ \varepsilon^{ij} E_{j}+\alpha^{ij}B_{j}, \\
    H^{i} & = & -M^{S}_{i} + \bar{\mu}^{ij}B_{j}-\alpha^{ji}E_{j},
\end{eqnarray}
where the permittivity and the inverse-permeability 3-d tensors were respectively defined by $\varepsilon^{ij}=\varepsilon_{0}( \delta^{ij}+\chi^{ij})$ and $\bar\mu^{ij}=\frac{1}{\mu_{0}}(  \delta^{ij}-\tilde{\chi}^{ij})$.

The constitutive relations describing a linear magnetoelectric medium can be presented using a covariant notation as follows \cite{hwang2024gravity,hwang2023maxwell}
\begin{eqnarray}
\label{D}
D^{\mu} & = & \varepsilon^{\mu\nu}E_{\nu}+\alpha^{\mu\nu}B_{\nu},\\
\label{H}
H^{\mu} & = & \bar\mu^{\mu \nu}B_{\nu}-\alpha^{\nu\mu}E_{\nu},
\end{eqnarray}
where $\varepsilon^{\mu\nu} = \delta^{\mu}_i\delta^{\nu}_j \varepsilon^{ij}$, $\bar\mu^{\mu\nu} = \delta^{\mu}_i\delta^{\nu}_j \bar\mu^{ij}$, and $\alpha^{\mu\nu} = \delta^{\mu}_i\delta^{\nu}_j \alpha^{ij}$. 
The electric and magnetic 4-vectors are defined in terms of the electromagnetic tensor $F_{\mu\nu}$ as $E_{\alpha}=F_{\alpha\beta}V^{\beta}$ and $cB_{\mu}=F^{*}_{\mu\beta}V^{\beta}$, respectively, with $V^\mu$ denoting the velocity field of the observer. 

The projector onto the 3-dimensional hypersurface ${\cal V}$ orthogonal to the observer's 4-velocity $V^\mu$ is given by $h_{\mu}{}^{\nu}=\delta_{\mu}{}^{\nu}+\tfrac{1}{c^2}V_{\mu}V^{\nu}$, with  $V^{\mu}V_{\mu}=-c^2$. 

The antisymmetric generalized tensor $\mathcal{G}_{\alpha\beta}$ that describes the electromagnetic field in a material medium can be written as
\begin{eqnarray}
    \nonumber
    \mathcal{G}^{\alpha\beta} &=& D^{\beta}V^{\alpha}-D^{\alpha}V^{\beta}-\tfrac{1}{c}\eta^{\alpha\beta}{}_{\lambda\mu}V^{\lambda}H^{\mu}\\
    \nonumber
    &= & V^{\alpha}\left(  \varepsilon^{\beta\nu}F_{\nu\mu}V^{\mu}+\tfrac{1}{c}\alpha^{\beta\nu}F^{*}_{\nu\mu}V^{\mu}\right)\\
    \nonumber
    & & -V^{\beta}\left(  \varepsilon^{\alpha\nu}F_{\nu\mu}V^{\mu}+\tfrac{1}{c}\alpha^{\alpha\nu}F^{*}_{\nu\mu}V^{\mu} \right)\\
    & &-\tfrac{1}{c} \eta^{\alpha\beta}{}_{\lambda\mu}V^{\lambda}\left(\tfrac{1}{c}  \bar\mu^{\mu \nu}F^{*}{}_{\nu\sigma}V^{\sigma}-\alpha^{\nu\mu}F_{\nu \sigma}V^{\sigma} \right).
    \label{G}
\end{eqnarray}
Therefore, in the absence of free sources of charge and current densities, the field equations are given by 
\begin{equation}
    \mathcal{G}^{\alpha\beta}{}_{,\beta}=0.
    \label{divG}
\end{equation}
To study wave solutions, the Hadamard method of field discontinuities is considered. In this framework, the electromagnetic fields are assumed to be continuous through the wavefront $\Sigma$ for any given instant of time $t$, while their first derivatives may present finite ``jumps" across the wavefront. Thus, it is imposed that $[F_{\mu\nu}]_{\Sigma}=0$ and $[\partial_{\lambda}F_{\mu\nu}]_{\Sigma}=e_{[\mu}k_{\nu]}k_{\lambda}$, \cite{2000PhLB..482..134D} where $k_\mu$ denotes the wave 4-vector, which is normal to the surface $\Sigma$, and $e_\mu$ denotes the polarization vector of the wave, associated with its electric field mode. 

The Harmard's method yields $\left[\mathcal{G}^{\alpha\beta}{}_{,\beta}\right]_\Sigma = 0$, and after some algebra, it follows that
\begin{eqnarray}
Z^{\alpha\rho}e_{\rho} =0,
\label{eveq}
\end{eqnarray}
where the effective Fresnel tensor $Z^{\alpha\rho}$ is defined by 
\begin{align}
    &Z^{\alpha\rho} = V^{\alpha}k_\beta \Big( V^{\mu}k_{\mu} \varepsilon^{\beta\rho}-\varepsilon^{\beta\nu}k_{\nu}V^{\rho}+\frac{1}{c}\alpha^{\beta\nu}\eta_{\nu\mu}{}^{\rho\delta}V^{\mu}k_{\delta}\Big)\nonumber \\
    &\qquad -V^{\beta}k_{\beta}\Big(V^{\mu}k_{\mu} \varepsilon^{\alpha \rho}-\varepsilon^{\alpha\nu}k_{\nu}V^{\rho}+\frac{1}{c}\alpha^{\alpha\nu}\eta_{\nu\mu}{}^{\rho\delta}V^{\mu}k_{\delta}\Big)\nonumber \\
    &\quad -\frac{1}{c}\eta^{\alpha\beta}{}_{\lambda\delta}k_{\beta}V^{\lambda}k_\gamma\Big(\frac{1}{c}\bar\mu^{\delta\nu}\eta_{\nu\mu}{}^{\rho\gamma}V^{\mu}- \alpha^{\rho\delta}V^{\gamma}  + \alpha^{\gamma\delta}V^{\rho}\Big).
     \nonumber
\end{align}
Non-trivial solutions for the wave modes require a singular Fresnel matrix, which can be found directly by means of the secular equation $\det(Z^{\alpha\rho})=0$. However, since the Fresnel tensor is orthogonal to the wave 4-vector ({\em i.e.}, it satisfies $Z^{\alpha\rho}k_{\rho}=0$ and $Z^{\alpha\rho}k_{\alpha}=0$), it suffices to look for solutions of this eigenvalue problem restricted to the 3-d space orthogonal to $k_{\nu}$. Hence, the secular equation reads $\det(Z^{\alpha\rho})=\tfrac{1}{6}[(Z_{1})^3-3Z_{1}Z_{2}+2Z_{3}]$, where $Z_{1}=Z^{\mu}{}_{\mu}$, $Z_{2}=Z^{\mu}{}_{\nu}Z^{\nu}{}_{\mu}$ and $Z_{3}=Z^{\alpha}{}_{\nu}Z^{\nu}{}_{\mu}Z^{\mu}{}_{\alpha}$ (see the appendix for further details). 

To solve this eigenvalue problem for the case of metamaterials, it is necessary to know the symmetries of the material under consideration. The point group of a crystalline system determines the directional variation of the optical properties of the material, and the magnetic point groups govern its magnetic properties \cite{authier2014international}.

To study wave propagation in a crystalline system in the presence of a background gravitational field it is considered the Schwarzschild metric for a central mass $M$, i.e.  $ds^2 =-f dt^{2}+f^{-1}dr^2+r^2 d\Omega^2 $, where $f\doteq f(r)=1-\tfrac{2GM}{c^2r}$, and the  2-sphere line element is given by $d\Omega^2=d\theta^2+\sin^2\theta\, d\phi^2$. Hereafter, the geometric mass $m = G M/c^2$ is used, and the wave 4-vector is assumed to be radially directed outwards, i.e., $k_{\mu}=(\tfrac{\omega}{c},q,0,0)$ with $q>0$.

\section{Optics in linear dielectrics in a gravitational field}
\label{optics_linear}
Hereafter, $V^\mu = c \delta^{\mu}_0$ is set, and the medium is chosen to be characterized by the following optical coefficients, expressed in a Cartesian coordinate system:

\begin{eqnarray}
    \varepsilon^{\mu\nu} &=& {\rm diag} (0,\varepsilon_{1},\varepsilon_{2},\varepsilon_{3}) \label{epsilon}, \\
    \bar\mu^{\mu\nu} &=& {\rm diag} (0,\tfrac{1}{\mu_{1}},\tfrac{1}{\mu_{2}},\tfrac{1}{\mu_{3}}) \label{mu},\\
    \alpha^{\mu\nu} &=& {\rm diag} (0,\alpha_{1},\alpha_{2},\alpha_{3}). \label{alpha}
\end{eqnarray}
Their representation in spherical coordinates can be obtained with the help of the basis-transformation matrix
\begin{equation}
    \Lambda^{\mu}{}_{\nu}=\left(
\begin{array}{cccc}
  1 & 0 & 0 & 0 \\
 0 &  {\scriptstyle \sin \theta \cos\phi} &{\scriptstyle \sin \theta \sin\phi} & {\scriptstyle \cos \theta }\\
 0 & \frac{\cos \theta \cos\phi}{r} & \frac{\cos \theta \sin\phi}{r} & -\frac{\sin \theta}{r} \\
 0 & -\frac{\sin\phi}{r \sin \theta} & \frac{\cos\phi}{r \sin \theta} & 0 \\
\end{array}
\right),
\label{TransfMat}
\end{equation}
where $\theta$ and $\phi$ are the Schwarzschild spherical coordinates. It is set $\phi=0$ for simplicity, yielding the above optical tensors to read 
\begin{equation*}
   \sigma^{\mu\nu}= \left(
\begin{array}{cccc}
0 & 0 & 0 & 0 \\
0 & {\scriptstyle \sigma_{1} \sin ^2\theta+\sigma_{3} \cos ^2\theta} & \frac{\sin \theta \cos \theta (\sigma_{1}-\sigma_{3})}{r} & 0 \\
 0 & \frac{\sin \theta \cos \theta (\sigma_{1}-\sigma_{3})}{r} & \frac{\sigma_{1} \cos ^2\theta+\sigma_{3} \sin ^2\theta}{r^2} & 0 \\
0 & 0 & 0 & \frac{\sigma_{2} \csc ^2\theta}{r^2} \\
\end{array}
\right)
\end{equation*}
where the kernel symbol $\sigma$ denotes any among $\varepsilon$, $\bar\mu$ or $\alpha$. That is to say, all three coefficient matrices explicitly display the same algebraic structure among their components. 

To obtain the wave-propagation solutions of the field equations (\ref{divG}), Eq.~(\ref{eveq}) must be solved. For the case of the Schwarzschild metric, with $\theta=\pi/2$ for simplicity, the effective Fresnel matrix is 
\begin{equation}
   Z_{\alpha}{}^{\beta}= \left( 
    \begin{array}{cccc}
        {\scriptstyle c^2\varepsilon_{2}f }&{\scriptstyle  -c \varepsilon_{2}f v }&{\scriptstyle  0 }&{\scriptstyle  0} \\
         {\scriptstyle \frac{c \varepsilon_{2} v}{f} }&{\scriptstyle  -\frac{\varepsilon_{2}v^2}{f} }&{\scriptstyle  0 }&{\scriptstyle  0}\\ 
        {\scriptstyle 0 }& {\scriptstyle 0 }& {\scriptstyle \frac{f^2}{\mu_{2}}-\varepsilon_{3}v^2 }& {\scriptstyle   (\alpha_{2}-\alpha_{3})fv}\\ 
        {\scriptstyle 0 }&{\scriptstyle  0 }&{\scriptstyle   (\alpha_{2}-\alpha_{3})fv} &{\scriptstyle  \frac{f^2}{\mu_{3}}-\varepsilon_{2}v^2 } \\
    \end{array}
   \right),
   \nonumber
\end{equation}
where $v= \omega/q$ is the phase speed. To calculate such speed $v$, it is suitable to diagonalize $Z_{\alpha}{}^{\beta}$, that is, expressing this matrix in the basis formed by its own eigenvectors. On such a basis, the elements of the diagonal matrix $Z'_{\alpha}{}^{\beta}$ coincide with its eigenvalues, which are $Z'_{0}{}^{0}=0$ and
\begin{equation}
\begin{aligned}
Z'_{1}{}^{1} &= \frac{\varepsilon_{2}\,\big(c^{2}f^{2}-v^{2}\big)}{f} , \\[6pt]
Z'_{2}{}^{2} &= \frac{(\mu_{2}+\mu_{3}) f^{2}
               - \mu_{2}\mu_{3} (\varepsilon_{2}+\varepsilon_{3})v^{2} - \sqrt{A} }
               {2\,\mu_{2}\mu_{3}} , \\[6pt]
Z'_{3}{}^{3} &= \frac{(\mu_{2}+\mu_{3})f^{2} 
               - \mu_{2}\mu_{3}(\varepsilon_{2}+\varepsilon_{3}) v^{2}+\sqrt{A}}
               {2\,\mu_{2}\mu_{3}} ,
\end{aligned}
\end{equation}
where it was defined $A= 4 \mu_{2}^2 \mu_{3}^2 (\alpha_{2}-\alpha_{3})^2 f^2v^2 +\left[ (\mu_{3}-\mu_{2})f^2+\mu_{2} \mu_{3}  (\varepsilon_{2}-\varepsilon_{3})v^2\right]^2$. 

From the imposition that both $Z'_{2}{}^{2}$ and $Z'_{3}{}^{3}$ are simultaneously zero, it follows that \(A\) vanishes. By solving this for $v$, one obtains
\begin{eqnarray}
    v_{\pm}^2 = \frac{a+b+\Gamma\pm\sqrt{(a+b+\Gamma)^2-4ab}}{2ab}\,f^2,
    \label{phasevel}
\end{eqnarray}
where $a=\varepsilon_{2}\mu_{3}$, $b=\varepsilon_{3}\mu_{2}$, and $\Gamma=\mu_{2}\mu_{3}\left(\alpha_{2}-\alpha_{3}\right)^2$. The refractive index of the medium is defined as $n_{\pm} = c/v_{\pm}$, where $c=1/\sqrt{\mu_{0}\varepsilon_{0}}$ is the speed of light in vacuum.  
In the limit $m \to 0$, Eq.~(\ref{phasevel}) recovers early results \cite{fuchs1965,de2023remarks}. In what concerns the eigenvalue $Z'_{1}{}^{1}$, its corresponding eigenvector coincides with the wave vector $k_\mu$, which is not a propagating mode (for further details, see the Appendix).

As it is well known, the effective refractive index is related to the bending of light phenomenon in the Schwarzschild spacetime through~\cite{ye2008gravitational,perlick2004gravitational}
\begin{equation*}
    n=\frac{2}{\sqrt{f}}\frac{1}{\sqrt{f}+1-\frac{m}{r}}.
\end{equation*}
Although this effect is not the same as the one discussed here, their magnitudes, when an expansion in weak fields is considered, can be compared. In fact, by suppressing the anisotropy of the optical coefficients in Eq.~(\ref{phasevel}), the corresponding refractive indices reduce to $1+\tfrac{2m}{r}$.  

\subsection{Estimates}

Suppose that the magnetic point group under consideration is $\bar{3}'m'$ \cite{rivera2009short, rivera1994definitions}. This symmetry determines that the permittivity, permeability, and magnetoelectric tensors take the form given in Eqs.~(\ref{epsilon}-\ref{alpha}), with  $\varepsilon_{1}=\varepsilon_{2}=\varepsilon_{\perp}$, $\varepsilon_{3}=\varepsilon_{\parallel}$, and $\mu_{1}=\mu_{2}=\mu_{\perp}$, $\mu_{3}=\mu_{\parallel}$. Note that the propagation in the optical-axis direction corresponds to $\theta=0$,  while the perpendicular propagation corresponds to $\theta=\pi/2$. 

The difference in the index of refraction is given by $\Delta n=n_{-}-n_{+}$. For antiferromagnets, it is estimated that $\mu\approx \mu_0$, and for single-phase magnetoelectrics, $\mu_0\alpha$ can be considered to be of the order of $10^{-12}{\rm s}\,{\rm m}^{-1}$. Hence, implementing an expansion of $\Delta n$ for $\Gamma \ll 1$ and $m \ll 1$, it is obtained that
\begin{eqnarray}
    & \Delta n =c(\sqrt{b}-\sqrt{a})+\frac{c \Gamma}{2(\sqrt{b}-\sqrt{a})}+
   \frac{2m c (\sqrt{b}-\sqrt{a})}{r} + \nonumber \\ 
   & +\frac{c m \Gamma }{r(\sqrt{b}-\sqrt{a})}+\mathcal{O}(m^2,\Gamma^2). 
   \label{refindexapp}
\end{eqnarray}
The different corrections to the index of refraction can be identified as follows
\begin{eqnarray*}
    &\Delta n_{0}&=c(\sqrt{b}-\sqrt{a}) \\
    &\Delta n_{\alpha}&= \frac{c \Gamma}{2(\sqrt{b}-\sqrt{a})}\\ 
    &\Delta n_{m}&=\frac{2 cm  (\sqrt{b}-\sqrt{a})}{r}\\
    &\Delta n_{m\alpha}&=\frac{c m \Gamma }{r(\sqrt{b}-\sqrt{a})}.
\end{eqnarray*}

To test our calculations, consider a chromia (Cr$_{2}$O$_{3}$), an antiferromagnet crystal whose magnetic point group is $\bar{3}'m'$ \cite{dzyaloshinskii1959magneto, abdullah2014structural}.  Following Dzyaloshinskii's formulation \cite{dzyaloshinskii1959magneto}, the $z$-axis of the Cartesian coordinate system is chosen to be parallel to the optical axis of the Cr$_{2}$O$_{3}$ system. Thus, the perpendicular and parallel components of $\varepsilon_{ij}$, $\mu_{ij}$, and $\alpha_{ij}$ optical coefficients are, respectively, perpendicular and parallel to the $z$-axis of the crystal. 
The crystalline medium is assumed to be effectively non-magnetic, with $\mu\approx\mu_{0}$. The magnetoelectric coefficients are approximated as $\mu_{0}\alpha_{\perp}=4.13\times 10^{-12} {\rm s}\,{\rm m}^{-1}$ and $\mu_{0}\alpha_{\parallel}=-0.52\times 10^{-12} {\rm s}\,{\rm m}^{-1}$ \cite{hehl2008relativistic}. In addition, using $\varepsilon_{\perp}/\varepsilon_{0}=11.9$ and $\varepsilon_{\parallel}/\varepsilon_{0}=13.3$ from the literature \cite{fang1963dielectric}.

Assuming the Sun is the source of gravity, for which $m= G M_{\odot}c^{-2}$, and the average radius of the Earth's orbit is $r=1.5\times 10^{11}{\rm m}$, it is obtained that the different contributions to the birefringence effect along the x-direction are
\begin{align}
    \Delta n_{0} &= 1.97\times 10^{-1}
    \nonumber\\
    \Delta n_{\alpha} &= 4.93\times 10^{-6} 
    \nonumber
\end{align}
\begin{align}
    \Delta n_{M_{\odot}} &= 7.82\times 10^{-9} 
    \nonumber\\
    \Delta n_{M_{\odot}\alpha} &= 1.95\times 10^{-13}
\nonumber
\end{align}

On the other hand, if Earth is taken as the source of gravity, where $m = G M_\oplus c^{-2}$, with $M_\oplus =6\times 10^{24}{\rm kg}$ and $r=6.378\times 10^{6}{\rm m}$, it yields
\begin{align}
     \Delta n_{0} & =   1.97\times 10^{-1} 
     \nonumber\\
      \Delta n_{\alpha} &=  4.93\times 10^{-6}  
      \nonumber\\
      \Delta n_{M_\oplus} & =  5.44\times 10^{-10} 
      \nonumber\\
      \Delta n_{M_\oplus\alpha}& =  1.36\times 10^{-14} 
      \nonumber
\end{align}

Once $\Delta n_{M_\oplus}$ and $\Delta n_{M_\oplus\alpha}$ are non-zero only due to the presence of the gravitational field, these results depend on the radial distance. Thus, experiments running at different altitudes would yield different outcomes. For instance, a variation in altitude of $50$km, which would be possible by using a high-altitude balloon, would lead to $\Delta n_{M_\oplus} = 5.40\times 10^{-10}$. 

\begin{widetext}
\section{Plasma}
\label{plasmasection}
Let us derive the dispersion relation for a plasma system in the presence of a background gravitational field. The description of warm plasma is typically discussed in two regimes, depending on the ratio of the electron mass to the ion mass. Since an ionized plasma is not being considered, the analysis focuses on the high-frequency regime. 
Considering a warm, non-ionized plasma with anisotropic optical properties, which includes finite temperature effects, the framework of fluid plasmas should be used \cite{swanson2020plasma,bittencourt2013fundamentals,zhao2017properties}. In Cartesian coordinates, the permittivity tensor that describes this system is given by 
 \begin{equation}
    \varepsilon^{\mu\nu}= \varepsilon_{0} \left(
\begin{array}{cccc}
 0 & 0 & 0 & 0 \\
 0 & 1-\frac{\omega_{p}^2 \tau (\theta )}{\tau (0) \omega ^2-\omega_{c}^2 \tau (\theta )} & \frac{i \omega_{c} \omega_{p}^2 \tau (\theta )}{\tau (0) \omega ^3-\omega  \omega_{c}^2 \tau (\theta )} & -\frac{c_{e}^2 q^2 \omega_{p}^2 \sin \theta \cos \theta}{\tau (0) \omega ^2-\omega_{c}^2 \tau (\theta )} \\
 0 & -\frac{i \omega_{c} \omega_{p}^2 \tau (\theta )}{\tau (0) \omega ^3-\omega  \omega_{c}^2 \tau (\theta )} & 1-\frac{\tau (0) \omega_{p}^2}{\tau (0) \omega ^2-\omega_{c}^2 \tau (\theta )} & -\frac{i c_{e}^2 q^2 \omega_{c} \omega_{p}^2 \sin \theta \cos \theta}{\tau (0) \omega ^3-\omega  \omega_{c}^2 \tau (\theta )} \\
 0 & -\frac{c_{e}^2 q^2 \omega_{p}^2 \sin \theta \cos \theta}{\tau (0) \omega ^2-\omega_{c}^2 \tau (\theta )} & \frac{i c_{e}^2 q^2 \omega_{c} \omega_{p}^2 \sin \theta \cos \theta}{\tau (0) \omega ^3-\omega  \omega_{c}^2 \tau (\theta )} & 1-\frac{\omega_{p}^2 \left(-c_{e}^2 q^2 \sin ^2\theta+\omega ^2-\omega_{c}^2\right)}{\tau (0) \omega ^2-\omega_{c}^2 \tau (\theta )} \\
\end{array}
\right),
\label{permit_plasmawarm}
    \end{equation}
\end{widetext}
where $\omega_{p}^2=N e^2/ (\varepsilon_{0} m)$ is the electron plasma frequency, $\omega_{c}= e B_{0}/m$ is the electron cyclotron frequency, and $c_{e}=\gamma  \kappa_{\rm B}  T/m$ is the sound speed for electrons. The function $\tau (\theta )=\omega ^2-c_{e}^2 q^2 \cos ^2\theta$ is defined for convenience. The permittivity tensor for a cold plasma can be immediately recovered by taking $c_{e}=0$ in Eq.~(\ref{permit_plasmawarm}). Additionally, setting $\bar{\mu}^{\mu\nu}=\mu^{-1}{\rm diag}(0,1,1,1)$ and $\alpha^{\mu\nu}=0$, is appropriate for plasma systems. Using the basis-transformation matrix, Eq.~(\ref{TransfMat}), with $\phi=0$, the permittivity tensor in Eq.~(\ref{permit_plasmawarm}) can be used to calculate the Fresnel tensor $Z^{\mu\nu}$. 
The corresponding dispersion relation is calculated with the help of the formula $Z_{1}^3-3Z_{1}Z_{2}+2Z_{3}=0$, that is:
\begin{widetext}
\begin{align}
\frac{6 \left(c^2 f^2 q^2-\omega^{2}\right)}{c^6 f \mu_{0}^3 [\tau (0) \omega^{2}-\omega_{c}^{2} \tau (\theta )]} \Big\{ &\omega_{c}^{2} \left(\omega^{2}-c^2 f^2 q^2\right) \left[\tau (\theta ) \left(c^2 f^2 q^2-\omega^{2}\right)+\omega_{p}^{2} \left(\omega^{2}-c^2 f^2 q^2 \cos ^2\theta \right)\right] \nonumber \\
&+\omega^{2} (\tau (0)-\omega_{p}^{2}) \left(c^2 f^2 q^2-\omega^{2}+\omega_{p}^{2}\right)^2\Big\}=0.
\label{disp_rel_warm}
\end{align}
\end{widetext}

It is worth noticing that Eq.~(\ref{disp_rel_warm}) contains the vacuum solution.  
To make this point clear, let us assume a cold plasma regime with  $\omega_{c}\rightarrow\infty$, for $\theta=\pi/2$. In this particular case, the dispersion relation reduces to $$\frac{6 \left(\omega ^2-c^2 f^2 q^2\right)^2 \left(c^2 f^2 q^2-\omega ^2+\omega_{p}^2\right)}{c^6 f \mu_{0}^3}=0$$ and the diagonalized Fresnel matrix reads 
$$\tfrac{1}{\mu_0 c^2}\,{\rm diag}(c^2 f^2 q^2-\omega ^2,c^2 f^2 q^2-\omega ^2 ,c^2 f^2 q^2-\omega ^2+\omega_{p}^2 ).$$ 
By solving the eigenvalue problem with $\omega^2 = c^2f^2q^2$, two zero eigenvalues associated to the eigenvectors $\{1,0,0\}$ and $\{0,1,0\}$ are obtained. The remaining eigenvector $\{0,0,1\}$ is associated to the eigenvalue $\lambda=\omega_{p}^2/(c^2\mu_{0})$, which indicates that the corresponding mode requires $\omega_{p}=0$, i.e., the absence of plasma. 
Therefore, hereafter, only the spatial part of the optical quantities will be considered. Remember that the presence of gravity is encoded in the function $f$.

It is worth exploring the warm and cold plasma regimes. By assuming $c_{e}$ is very small compared to the $v_{ph}$, Eq. (\ref{permit_plasmawarm}) can be conveniently expanded in powers of $\xi = (c_{e}/v_{ph})^2$, which, up to second order, results in
     \begin{eqnarray}
    &\frac{\varepsilon^{ij}}{\varepsilon_{0}}  =  \left(
\begin{array}{ccc}
 1-\frac{\omega_{p}^2 }{\omega ^2-\omega_{c}^2 } & \frac{i \omega_{c} \omega_{p}^2 }{\omega ^3-\omega  \omega_{c}^2 } & 0 \\
 -\frac{i \omega_{c} \omega_{p}^2 }{\omega ^3-\omega  \omega_{c}^2 } & 1-\frac{\omega_{p}^2}{\omega ^2-\omega_{c}^2 } & 0 \\
 0& 0& 1-\frac{\omega_{p}^2 }{\omega ^2} \\
\end{array}
\right) &\nonumber\\
&\scriptstyle{-\xi \omega_p^2 \sin^2\theta}  \left(
\begin{array}{ccc}
     \frac{\omega^2}{(\omega^2-\omega_{c}^2)^2}  & -i \frac{\omega \omega_{c}}{(\omega^2-\omega_{c}^2)^2}  & \frac{\cot \theta}{\omega^2-\omega_{c}^2} \\
    i\frac{\omega \omega_{c}}{(\omega^2-\omega_{c}^2)^2}  & \frac{\omega_{c}^2}{(\omega^2-\omega_{c}^2)^2}  & i\frac{\omega_{c}\cot \theta}{\omega(\omega^2-\omega_{c}^2)}  \\
    \frac{\cot\theta }{\omega^2-\omega_{c}^2} & -i\frac{\omega_{c}\cot\theta}{\omega(\omega^2-\omega_{c}^2)}  & \frac{\cot^2\theta}{\omega^2}  \\
\end{array}
\right).&\hspace{1.2em}
\label{permitseries}
    \end{eqnarray}

Looking for the eigenvalues at first order in $\xi$, the terms in the diagonalized matrix $\varepsilon^{\mu\nu}_{D}$ are
\begin{eqnarray*}
    \frac{\varepsilon^{11}_{D}}{\varepsilon_{0}} & = &   
     1-\frac{\omega_{p}^2}{\omega^2}\left(1+\frac{c_{e}^2}{v_{ph}^2} \cos^2\theta \right), \\
     \frac{\varepsilon^{22}_{D}}{\varepsilon_{0}} & = &  1-\frac{\omega_{p}^2}{\omega (\omega+\omega_{c})}\left[1+\frac{c_{e}^2}{v_{ph}^2}\frac{\omega \sin^2\theta}{2(\omega+\omega_{c})} \right], \\
     \frac{\varepsilon^{33}_{D}}{\varepsilon_{0}} & = &  
      1-\frac{\omega_{p}^2}{\omega (\omega-\omega_{c})} \left[1+\frac{c_{e}^2}{v_{ph}^2}\frac{\omega \sin^2\theta}{2(\omega-\omega_{c})} \right] ,
\end{eqnarray*}
where the second term in each component of $\varepsilon^{ij}_{D}$ corresponds to the cold-plasma contribution, and the last term is its thermal correction.

Let us investigate a possible realization of these results. The correction due to the spacetime curvature in a Tokamak device on Earth is usually neglected as $2m/r$ for the Earth is of the order of $10^{-9}$. Suppose that such a device is operated by a fluid with plasma frequency $\omega_{p}=5.7\times10^{11} \mbox{rad s}^{-1}$, under the action of a magnetic field of $3\mbox{T}$, its electron cyclotron frequency is $\omega_{c}=5.3\times 10^{11} \mbox{rad s}^{-1}$ and a temperature of $T=10^{8}$K, given the sound speed for electrons as $c_{e}=6.8\times10^{7}{\rm m\,s}^{-1}$ \cite{goedbloed2019magnetohydrodynamics}.  Then, the phase velocity must be $v_{ph}>0.23 c$.
Moreover, for $\theta=0$, when the phase velocities are calculated for the warm plasma, they coincide with those considering a cold plasma, i.e., 
\begin{align}
(v_{\pm}^2)_{\rm{cold}}=(v_{\pm}^2)_{{\rm warm}}=\frac{c^2 \omega(\omega\pm\omega_{c})}{\omega^2\pm\omega \omega_{c}-\omega_{p}^2},
\nonumber
\end{align}
which shows that, in this particular case, both frameworks are interchangeable.

For an angle $\theta=\pi/2$ the dispersion relation expressed in Eq. (\ref{disp_rel_warm}) reduces to 
\begin{align} 
&c^2 c_{e}^2 q^4+q^2 \left[c^2 (-\omega^{2}+\omega_{c}^{2}+\omega_{p}^{2})+c_{e}^2 (\omega_{p}^{2}-\omega^{2})\right]\nonumber  \\
& -\omega^{2} \omega_{c}^{2}+(\omega^{2}-\omega_{p}^{2})^2=0.
\end{align}
For the cases where $c_{e} \ll c$, the second term in the square brackets can be neglected \cite{bittencourt2013fundamentals}. 
However, for the case of a Tokamak machine \cite{goedbloed2019magnetohydrodynamics}, $c_{e} \approx 0.23 c$, and this term might be significant. 

To solve the dispersion relation for any angle, a numerical analysis is needed. As the focus of this paper is to consider the influence of gravity, such analysis will be postponed to a future project.

\subsection{Magnetosphere of a neutron star}
An interesting application of the above-discussed formalism occurs for the magnetosphere of a neutron star. Suppose it has a radius of $r=10\, {\rm km}$ and its mass $M_{NS}$ ranges between 1.5 and 2.5 solar masses $M_{\odot}$, such that $f(r)$ ranges between $0.56$ and $0.27$.
Neutron stars typically have a magnetic field of the order of $B\sim 10^{8} \,{\rm T}$, associated with a cyclotron frequency of $\omega_{c}\sim 10^{20} \,{\rm rad}\, {\rm s}^{-1}$. The temperature in the magnetosphere is of the order of $T=1.6\times 10^6\,{\rm K}$, leading to a sound velocity of $c_{e}=8.53\times 10^6\,{\rm m \, s}^{-1}$ \cite{michel1982theory, petri2016theory}. Assuming a plasma frequency $\omega_{p}=1.78\times 10^{10}\,{\rm rad \, s}^{-1}$ \cite{petri2016theory}, the regime of $\omega \ll \omega_{c}$ applies.

Up to linear order in $\xi=\tfrac{c_e}{v_{ph}}$, the permittivity tensor reads
 \begin{eqnarray*}
   \frac{\varepsilon^{ij}}{\varepsilon_{0}}= \left(
\begin{array}{ccc}
 1 & 0 & 0 \\
  0 & 1 & 0 \\
  0 & 0 & 1-\frac{\omega_{p}^2 }{\omega ^2} \\
\end{array}
\right)
-\frac{c_{e}^2}{v_{ph}^2} \left(
\begin{array}{ccc}
     0 & 0 & 0 \\
     0 & 0 & 0 \\
     0 & 0 & \frac{\omega_{p}^2\cos^2\theta}{\omega^2}  \\
\end{array}
\right).
    \end{eqnarray*}

Notice that the system is naturally anisotropic, presenting an optical axis in the z-direction, anticipating birefringence phenomena. The only non-trivial component is 
\begin{equation}
    \frac{\varepsilon^{33}}{\varepsilon_0}=1-\frac{\omega_{p}^2}{\omega^2}\left(1+\cos^2 \theta \frac{c_{e}^2}{v_{ph}^2} \right),
\end{equation}
where the last term, which is the small thermal contribution, decreases as either the phase velocity or the angle $\theta$ increases. Writing this equation as
\begin{equation}
    \frac{\varepsilon^{33}}{\varepsilon_0}=1-\Delta\varepsilon_{P}-\Delta \varepsilon_{PW},
\end{equation}
where, for $\theta=0$, $\Delta\varepsilon_{P}\sim 3.16\times 10^{-2}$ is the modification for the plasma with respect to the vacuum, and $\Delta\varepsilon_{PW}\sim 1.02\times 10^{-2}$ is the correction to the plasma due to the temperature. Considering these corrections, the warm plasma regime seems to be a reliable approach.

To compare the phase velocity with the value of $c_{e}$ for any $\theta$, the dispersion relation for a warm plasma in the limit $\omega_{c}\rightarrow\infty$ is considered:
\begin{eqnarray}
\frac{6 \left(\omega^{2}-c^2 f^2 q^2\right)^2}{c^6 f \mu_{0}^3 \left(\omega^{2}-c_{e}^2 q^2 \cos ^2\theta\right)}\Big[\omega^{2} \left(c^2 f^2 q^2-\omega^{2}+\omega_{p}^{2}\right)
\nonumber \\ 
+ q^2 \cos ^2\theta \left(c_{e}^2 \omega^{2}-c^2 f^2 \left(c_{e}^2 q^2+\omega_{p}^{2}\right)\right)\Big]=0.
\label{disp_relat_limit}
\end{eqnarray}
The phase velocity can be obtained as $v_\pm = \omega/q_\pm$, with
\begin{align}
   &q_{\pm}=\frac{1}{2} \left[\frac{\omega^2}{f^2c^2}+ \frac{\omega^{2}-\omega_{p}^{2}\cos^2\theta}{c_{e}^2\cos ^2\theta}  \right.
   \nonumber \\
   &\pm \left. \sqrt{\left(\frac{\omega^2}{f^2c^2}+ \frac{\omega^{2}-\omega_{p}^{2}\cos^2\theta}{c_{e}^2\cos ^2\theta} \right)^2-4 \frac{\omega^2 (\omega^{2}-\omega_{p}^{2})}{f^2c^2c_{e}^2\cos ^2\theta}}\; \right].
   \nonumber
\end{align}

\begin{figure}[ht]
    \centering
    \includegraphics[width=\linewidth]{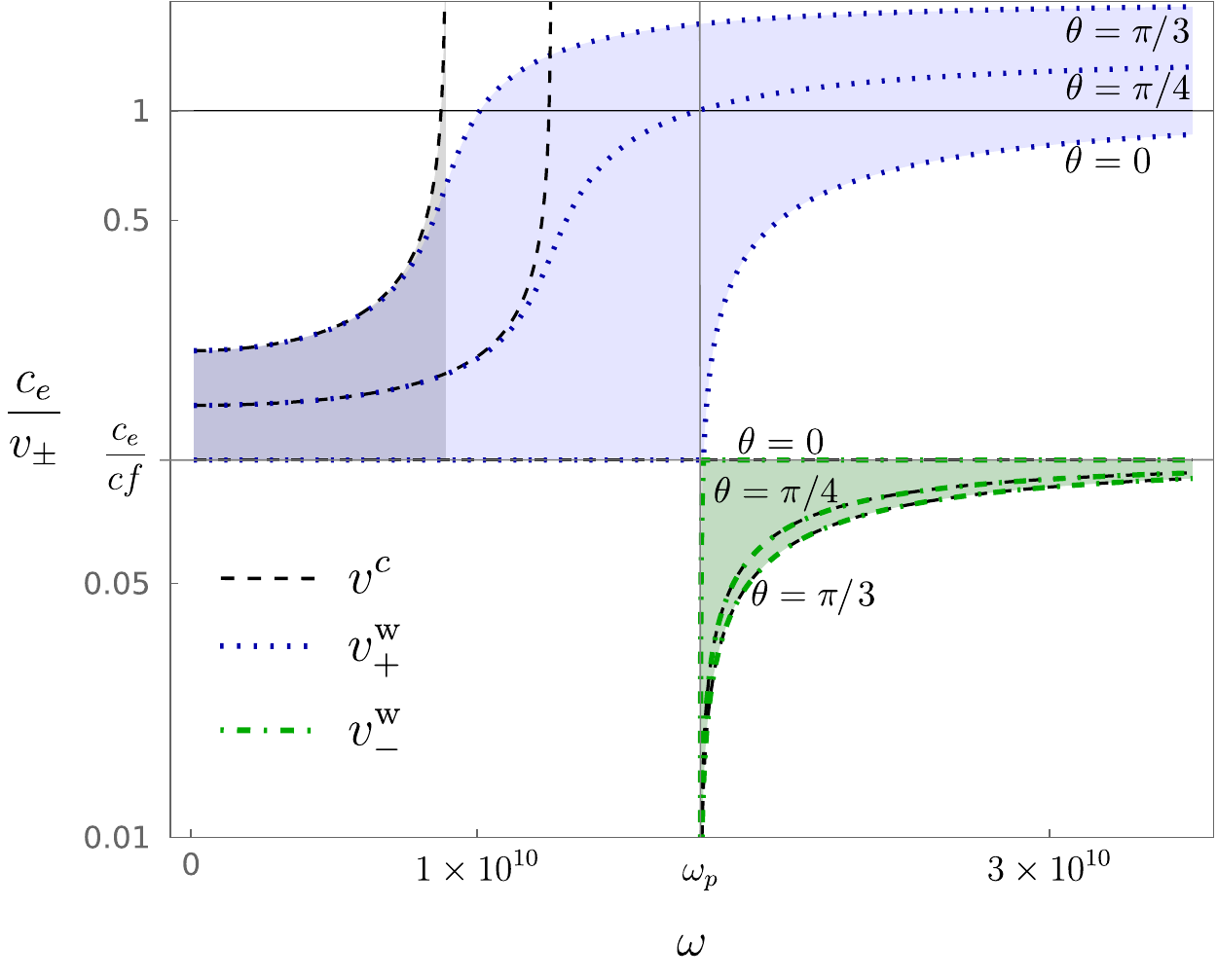}
    \caption{Ratio between the thermal velocity $c_{e}$ and the phase velocity $v_{\pm}$ with respect to the frequency $\omega$. The dotted curve indicates the solution for the cold plasma case $v^{c}$, the dashed curve is the solution for the warm positive solution $v^{\rm w}_{+}$, while the dot-dashed curve is the solution for the warm negative solution $v^{\rm w}_{-}$. The shaded area indicates how the curves evolve from $\theta=0$ to $\theta=\pi/4$. The parameter $c_{e}/(cf)=0.109$. For improved visualization, the vertical axis is shown on a logarithmic scale. }
    \label{fig:tempvelVSphasevel}
\end{figure}

The ratio between the wave phase velocity and the thermal velocity is explored in Fig.~\ref{fig:tempvelVSphasevel}. 
Note that for small angles, $\theta\approx0$, the warm-plasma solutions $v^{\rm w}_{\pm}$ lie within the validity regime. However, as $\omega$ and $\theta$ increase, $v^{\rm w}_{+}\rightarrow c_{e}$. The shaded area in the figure indicates how the curves evolve from angle $\theta=0$, for which the warm and cold plasma solutions coincide, up to the angle $\theta=\pi/4$. The presence of gravity, contained in $f$, affects the final value of the asymptotes. As $f\rightarrow 1$, the asymptote tends to zero.  

\begin{figure}
    \centering
    \includegraphics[width=1\linewidth]{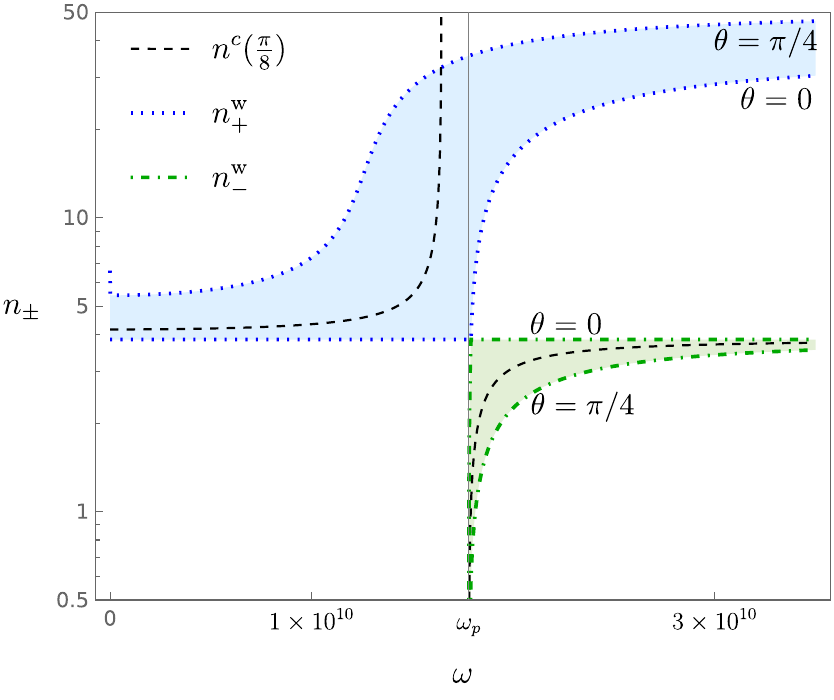}
    \caption{The indices of refraction for the warm and cold plasma frameworks are plotted with respect to the frequency $\omega$. The phase velocities $v^{\rm w}_{+}$ (blue shaded area), $v^{\rm w}_{-}$ (green shaded area) are plotted in a range of $\theta=(0-\pi/4)$. The refractive index for the cold plasma framework $n^{c}$ (dashed line) is plotted for an angle $\theta=\pi/8$ for the sake of comparison. Plasma frequency is $\omega_{p}=1.7\times 10^{10}\,{\rm rad \, s}^{-1}$ and $f=0.26$. The vertical axis is shown on a logarithmic scale.}
    \label{fig:comparativeCWindex}
\end{figure}

To further illustrate the difference between the cold and warm plasma frameworks, in Fig.~\ref{fig:comparativeCWindex}, the refractive indices given by the two models are compared. The choice $\theta=\pi/8$ illustrates that the refractive index for the cold plasma regime belongs to the range already established by the warm plasma solution. It is important to highlight the behavior of the refractive index as $\omega\rightarrow \omega_{p}$. In the cold plasma framework, the squared phase velocity is $v_{c}^2=c^2\left(\omega^2-\omega_{p}^2\cos^2\theta\right)/\left(\omega^2-\omega_{p}^2 \right)$. Therefore, at the frequency $\omega=\omega_{p}$ there is no propagation~\cite{bittencourt2013fundamentals}. 

\begin{figure}[ht]
    \centering
    \includegraphics[width=\linewidth]{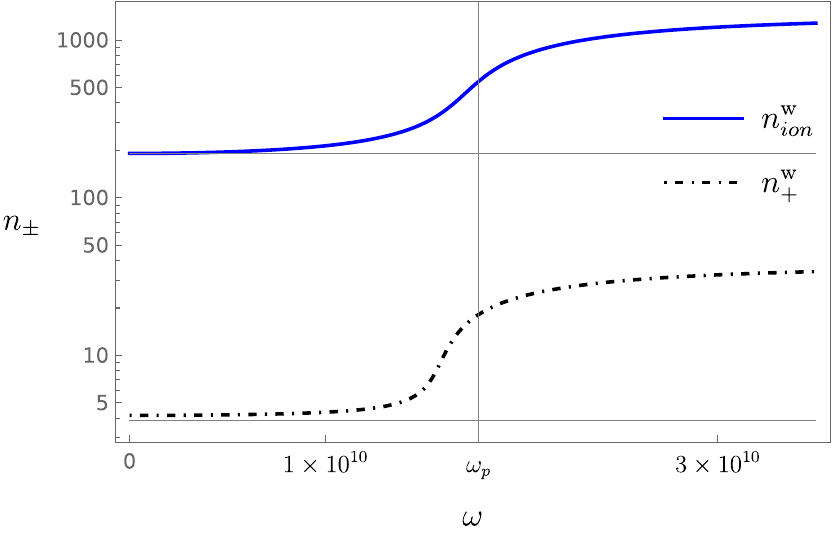}
    \caption{Refractive indices for the fully ionized magnetoplasma (thick line) and for the warm non-ionized magnetoplasma (dotted line). The lines indicating the asymptotes correspond to the values $f=0.26$ and $c/c_{e}=35.18$. Electron and ion plasma frequency are $\omega_{p}=1.7\times 10^{10}\,{\rm rad \, s}^{-1}$. The vertical axis is shown on a logarithmic scale.}
    \label{fig:NSion_warm}
\end{figure}
In the warm plasma case, only one mode propagates, whose phase velocity at $\omega=\omega_{p}$ is 
\begin{equation}
    v_{+}^2 =\frac{c^2 f^2 c_{e}^2 \cos^2\theta}{c^2 f^2+\left(c_{e}^2-c^2f^2\right)\cos^2 \theta}.
    \label{phasevel_warmNS}
\end{equation}
This existing mode at $\omega=\omega_p$ may be compared to the case of an ionized warm plasma, as examined in Fig.~\ref{fig:NSion_warm}.

In the case of a fully ionized warm magnetoplasma with ion-plasma frequency $\omega_{pi}$ and  ion-cyclotron frequency $\omega_{ci}$, the dispersion relation for $\theta=0$ in the limit $\omega_{c}\rightarrow\infty$ and $\omega_{ci}\rightarrow\infty$, and in the absence of gravity, is given by
\begin{align}
    &c_{e}^2c_{i}^2q^4+q^2\Big[\omega_{p}^2 c_{i}^2+\omega_{pi}^2 c_{e}^2-\omega^2 (c_{e}^2+c_{i}^2) \Big]\\
    &+\omega^2 (\omega^2-\omega_{p}^2-\omega_{pi}^2)=0.
    \nonumber
\end{align}
This is a biquadratic equation whose two solutions correspond to the ion-plasma ($v_{+}$) and electron-plasma waves ($v_{-}$)~\cite {bittencourt2013fundamentals}. 
The solution for the ion-plasma wave is displayed in Fig.~\ref{fig:NSion_warm} (solid curve), together with the solution for non-ionized warm plasma, described by Eq.~(\ref{phasevel_warmNS}). 
In this figure, it was assumed that the presence of the ions does not modify the electron plasma density and temperature, and that $\omega_{pi}$ coincides with $\omega_{p}$. 
Note that, while the shape of these curves is similar, the intensities differ greatly. 

\begin{figure}[t]
    \centering
    \includegraphics[width=\linewidth]{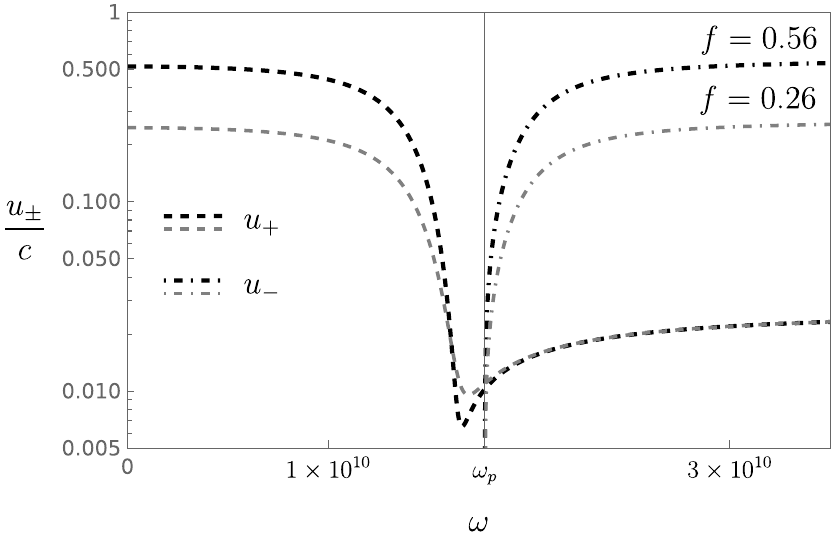}
    \caption{Group velocity of a light ray propagating in the magnetosphere of a neutron star for different values of the parameter $f$, with $M_{NS}=(1.5-2.5)M_{\odot}$. The parameters are chosen as $\omega_p=1.78\times 10^{10}{\rm rad\,s}^{-1}$, $\theta=\pi/8$ and $c_e/c=0.028$. The vertical axis is shown on a logarithmic scale.}
    \label{fig:NS_groupvel}
\end{figure}
\begin{figure}[t]
    \centering
    \includegraphics[width=\linewidth]{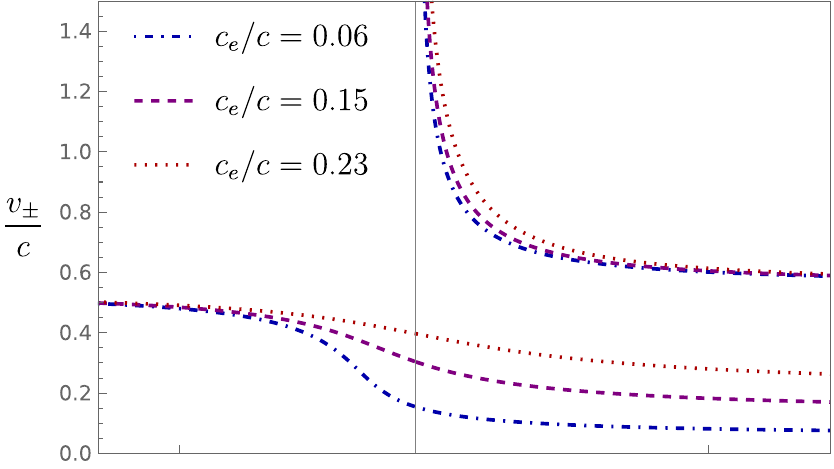}
    \includegraphics[width=\linewidth]{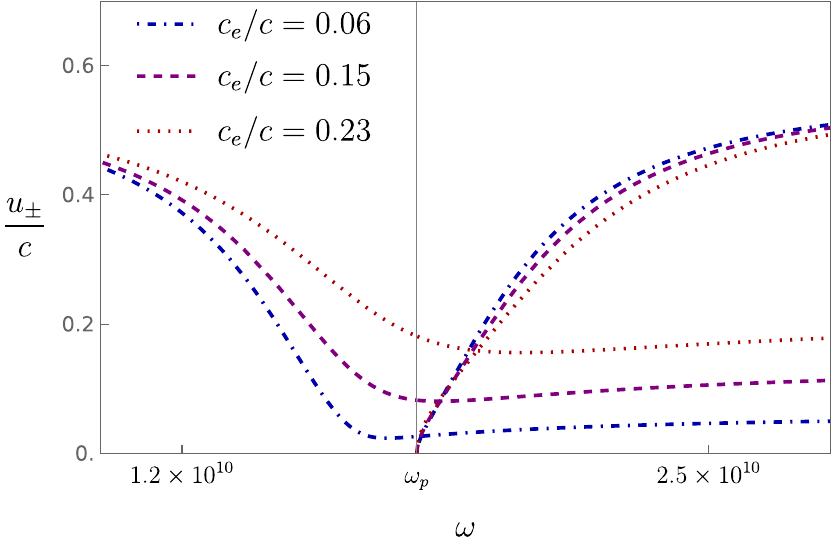}
    \caption{Phase and group velocity of a light ray propagating in the magnetosphere of a neutron star for a range of temperatures from $T=8.7\times10^{6}-1.4\times 10^{8}$K, which gives values of $c_{e}=2\times 10^7-8\times10^7\,{\rm m\,s}^{-1}$. It is set the value  $f=0.56$ at an angle $\theta=\pi/8$.}
    \label{fig:NS_temp}
\end{figure}

Finally, it is worth examining the behavior of the group velocity $u = d\omega/dq$, depicted in Figs.~\ref{fig:NS_groupvel} and \ref{fig:NS_temp}. Considering neutron-star masses ranging from  $M_{NS}=1.5M_{\odot}$ and $M_{NS}=2.5M_{\odot}$, for which $f=0.56$ and $f=0.27$, respectively. At the frequency $\omega=\omega_p$, $u_{-}$ vanishes for any $f$. However, $u_+/c=0.01234$ for $f=0.27$, and $u_+/c= 0.01231$ for $f=0.56$; the difference lies in the fifth decimal digit. For $\omega \neq \omega_p$, the group velocities for different values of $f$ are distinct from each other. Moreover, it is appreciated that for $\omega>\omega_p$ there is also birefringence. 
The birefringence effect can be seen in more detail in Fig.~\ref{fig:NS_temp}. 
Taking the limit $\omega\rightarrow\infty$, it results that $u_- = f$ and $u_+ = (c/c_e)\, f^2 \sec \theta$.
Additionally, the influence of the thermal velocity $c_{e}$ is also depicted. 

As depicted in Fig. \ref{fig:NS_groupvel}, the existence of two distinct propagation modes gives rise to birefringence whenever $\omega>\omega_p$. In this regime, both the $+$ and $-$ modes propagate with different polarization eigenvectors and generally different group velocities. Hence, an optical pulse received from this system would exhibit measurable patterns such as distinct arrival times and polarization-dependent pulse broadening. On the other hand,  for frequencies below the plasma frequency ($\omega<\omega_p$), only the $+$ mode propagates. Consequently,  birefringence disappears below the plasma frequency. This birefringence transition may offer a valuable fingerprint of the plasma environment and its coupling to gravity, particularly in the vicinity of compact astrophysical objects where both effects are expected to be significant.

\section{Final remarks}
\label{final}
The complete Fresnel tensor in the covariant formalism employed in this work allows us to incorporate the effects of gravity through the contraction of indices, as all tensor contractions are performed with the background metric. By choosing the background metric to be the Schwarzschild metric, even simple examples of non-trivial media, such as magnetoelectric crystals and magnetized plasmas, acquire gravitational contributions that are encoded in terms of the function $f=1-2m/r$. 
In Earth-based systems, as in the case of the magnetoelectric crystal considered in the text, the gravitational contribution is small, with refractive index differences of the order $\Delta n_{M_{\odot}}\sim 10^{-9}-10^{-10}$. However, optics laboratories using high-sensitivity interferometric techniques \cite{2025OptCo.58631829L} currently reach $\Delta n\sim10^{-11}$.

The gravitational effect becomes significantly larger in astrophysical environments near compact objects. In the magnetosphere of a neutron star with a strong magnetic field, such as a pulsar or a magnetar, wave propagation depends on the plasma environment and the star's gravitational field. The waves are generated within the plasma due to the spiral motion of electrons and positrons around magnetic field lines and are emitted in a tightly collimated cone guided along those lines. As these waves traverse the magnetosphere, they undergo several optical effects. In particular, as examined in this work, birefringence may occur due to the non-trivial constitutive relation of the plasma environment, and its magnitude will also depend on the gravitational field in the vicinity of the neutron star. In this way, the interplay of the anisotropic media with the gravitational field shapes the wave signals that could be detected from these extreme stars.
Although current instrumental precision could achieve such measurements, establishing observational signatures that allow these effects to be tested in realistic astrophysical scenarios remains challenging.

As the waves propagate throughout the plasma, the birefringence effect will split the rays as they emerge from the star environment. Thus, an observer at infinity will receive both wave modes at slightly different angles. More importantly, along the propagation through the plasma, the birefringence effect will lead to a time delay between the two light ray solutions. 

The results suggest that the propagation effects discussed in this work may be useful for obtaining observable signatures of gravity in anisotropic media. In particular, measurements of phase shifts, time delays, polarization, and angular splitting in radiation emitted by pulsars and magnetars could provide valuable information about the plasma constitutive properties and the spacetime geometry surrounding these objects. 
Recent studies in nonlinear electrodynamics report a non-negligible systematic correction to the time delay relative to the NICER mission results \cite{porto2026nonlinear}.

\begin{acknowledgments}
E. G. H. acknowledges CNPq (Conselho Nacional de Desenvolvimento Científico e Tecnológico) under Grant No. 151974/2024-1. V.~A.~D.~L. is supported in part by the Brazilian research agency CNPq under Grant No. 302492/2022-4. 
\end{acknowledgments}

\section*{Author Declarations}
All authors contributed equally to this work. The authors declare no conflicts of interest. The data that support the findings of this article are not publicly available. The data are available from the authors upon reasonable request.

\appendix
\section{The eigenvalues of the Fresnel matrix}
\label{setoffigs}
\setcounter{equation}{0}
\renewcommand\theequation{A\arabic{equation}}
\setcounter{figure}{0}
\renewcommand\thefigure{A\arabic{figure}}

The Fresnel equation $Z_{\mu}{}^{\nu}e_{\nu}=0$ determines the physics of the propagating modes, and non-zero eigenvectors $e_{\nu}=0$ satisfying this equation require $\det(Z_ {\mu}{}^{\nu})=0$. However, not all such eigenvectors (with eigenvalue $0$) are admissible as a polarization mode. Indeed, it is known in advance that the wave vector $k_{\nu}$ satisfies $Z_{\mu}{}^{\nu}k_{\nu}=0$, but $k_{\nu}$ cannot be taken as a polarization mode. Because the wave vector must have a non-zero time component, while physically meaningful $e_{\nu}$ are space-like vectors with vanishing time component, when expressed in terms of an orthonormal basis of spacetime vectors with Lorentz signature. Therefore, Fresnel matrix $[Z_{\mu}{}^{\nu}]$ lies in the 3-dimensional hyperplane orthogonal to the wave vector $k_{\nu}$. By considering a new orthogonal basis $(k_{\nu},b_{\nu}^{(1)},b_{\nu}^{(2)},b_{\nu}^{(3)})$ with $k$ in the basis, it follows that the Fresnel matrix is now expressed as $[Z_{\mu}{}^\nu]=0\oplus [Z_{i}]{}^{j}$, where the 3-dimensional matrix $Z_{i}{}^{j}$ is completely decomposable in terms of the three vectors $b_{\nu}^{(1)},b_{\nu}^{(2)},b_{\nu}^{(3)}$. Thus, the Fresnel equation could alternatively be written as $Z_{i}{}^{j}e_{j}=0$, with solutions $\bar{e}_{i}=\tilde{e}_{(1)}b_{i}^{(1)}+\tilde{e}_{(2)}b_{i}^{(2)}+\tilde{e}_{(3)}b_{i}^{(3)}$ decomposable in this new basis. This form of the Fresnel equation admits non-zero 3-dimensional eigenvectors $\tilde{e}_{i}=\bar{e}_{i}\neq 0$ if, and only if, the 3-dimensional matrix $[Z_{\mu}{}^{\nu}]$ has a vanishing 3-dimensional determinant $\det_{(3)}(Z_{i}{}^{j})=0$. But the 3-dimensional determinant $\det_{(3)}(A_{i}{}^{j})$ of any 3-dimensional matrix $A_{i}{}^{j}$ can be expanded as $\det_{(3)}(A_{i}{}^{j})=\frac{1}{6}(A_{1}^3-3A_{1}A_{2}+2A_{3})$, where $A_{1}=A_{i}{}^{i}$, $A_{2}=A_{i}{}^{j}A_{j}{}^{i}$ and $A_{3}=A_{i}{}^{j}A_{j}{}^{k}A_{k}{}^{i}$ are the traces of powers of $[A_{i}{}^{j}]$. The scalar equation then reads $\tilde{Z}_{1}^3-3\tilde{Z}_{1}\tilde{Z}_{2}+2\tilde{Z}_{3}=0$, where $\tilde{Z}_{1}=Z_{i}{}^{i}$, $\tilde{Z}_{2}=Z_{i}{}^{j}Z_{j}{}^{i}$, $\tilde{Z}_{3}=Z_{i}{}^{j}Z_{j}{}^{k}Z_{k}{}^{i}$ are the traces of powers of the 3-dimensional matrix $[Z_{i}{}^{j}]$. Note that $Z_{1}=\tilde{Z}_{1}$, $Z_{2}=\tilde{Z}_{2}$, $Z_{3}=\tilde{Z}_{3}$ in this new basis where $Z_{1}=Z_{\mu}{}^{\mu}$, $Z_{2}=Z_{\mu}{}^{\nu}Z_{\nu}{}^{\mu}$, $Z_{3}=Z_{\mu}{}^{\nu}Z_{\nu}{}^{\lambda}Z_{\lambda}{}^{\mu}$ are the traces of powers of the 4-dimensional matrix $[Z_{\mu}{}^{\nu}]$. This means that the secular equation can be equivalently written as $Z_{1}^3-3Z_{1}Z_{2}+2Z_{3}=0$. But this last equation is independent of the basis of vectors chosen, since the trace of a matrix is invariant with respect to any covariant change of the basis of vectors that span that matrix. 

Let us return to the original Lorentzian basis of vectors $\{T_{\mu},X_{\mu}^{(1)},X_{\mu}^{(2)},X_{\mu}^{(3)}\}$, since the eigenvector is $[e_{\nu}]=0\oplus[e_{i}]$ orthogonal to the time-like vector $T_{\mu}$, where $$ e_{i}=\sum_{\mu=0}^{3}\sum_{j=1}^{3}\frac{\partial b_{\mu}^{(j)}}{\partial x_{\mu}^{(i)}}\tilde{e}_{j}$$
are the components of $\tilde{e}_{j}$ transformed by the change of the basis. Thus, any time-component of the Fresnel matrix $[Z_{\mu}{}^{\nu}]$ will drop out upon contraction with $e_{\nu}$. As consequence, the matrix $[Z_{\mu}{}^{\nu}]$ can be projected in the 3-space orthogonal to $T_{\mu}$ by simply dropping out any of its time-components (which are not all zero), in the form $\bar{Z}_{\mu}{}^{\nu}=\sum_{i,j=1}^{3}\delta_{\mu}^{i}\delta_{j}^{\nu}Z_{i}{}^{j}$. Both $Z_{\mu}{}^{\nu}$ and $\bar{Z}_{\mu}{}^{\nu}$ determine the same eigenmodes $e_{\nu}$ with vanishing eigenvalue, which implies that $[\bar{Z}_{\mu}{}^{\nu}]=0\oplus[\bar{Z}_{i}{}^{j}]$ where $[\bar{Z}_{i}{}^{j}]$ is a 3-dimensional matrix with vanishing 3-dimensional determinant $\det_{(3)}(\bar{Z}_{i}{}^{j})=0$. As above, this is equivalently written as $\bar{Z}_{1}^3-2\bar{Z}_{1}\bar{Z}_{2}+2\bar{Z}_{3}=0$, where $\bar{Z}_{1}=\bar{Z}_{i}{}^{i}$, $\bar{Z}_{2}=\bar{Z}_{i}{}^{j}\bar{Z}_{j}{}^{i}$, $\bar{Z}_{3}=\bar{Z}_{i}{}^{j}\bar{Z}_{j}{}^{k}\bar{Z}_{k}{}^{i}$ are the traces of either powers of the 3-dimensional matrix $[\bar{Z}_{i}{}^{j}]$ or powers of the 4-dimensional matrix $[Z_{\mu}{}^{\nu}]$. As a result, the same eigenvalue equation $$Z_{1}^3-3Z_{1}Z_{2}+2Z_{3}=0$$ is obtained, yielding the same results for the modes and their velocities, where the matrix $[Z_{\mu}{}^{\nu}]$ can be interpreted either as the Fresnel matrix or the 3-space projection of the Fresnel matrix. 

\bibliography{ref}

\end{document}